\documentclass[a4paper]{jpconf}
\usepackage{graphicx}
\begin{document}
\title{A Powerful New Energy Density Functional}

\author{A.~W.~Thomas$^1$, P.~A.~M.~Guichon$^2$, J.~Leong$^1$, K.~L.~Martinez-Paglinawan$^3$ and J.~R. Stone$^4$}

\address{$^1$CSSM and ARC Centre of Excellence for Dark Matter Particle Physics,\\ 
Department of Physics, University of Adelaide SA 5005, Australia \\
$^2$IRFU-CEA, Universit\'e Paris-Saclay, F91191 Gif sur Yvette, France \\
$^3$Department of Physics, Silliman University, Dumaguete City, Negros Oriental, Philippines \\
$^4$Department of Physics (Astro), University of Oxford, OX1 3RH United Kingdom \, and \\
Department of Physics and Astronomy, University of Tennessee, TN 37996 USA}


\begin{abstract}
We describe the most recent energy density functional derived within the quark-meson coupling model. Although fit to the binding energies and charge radii of just seventy magic nuclei, the accuracy with which it reproduces nuclear properties across the entire periodic table is outstanding. As well as outlining a number of those results, we present an argument explaining why having a physically motivated model with a small number of parameters is especially desirable as one seeks to make predictions in new regions of $N$ and $Z$. As an example we show the predictions for known super-heavy nuclei that were not included in the fit. 
\end{abstract}

\section{Introduction}
Here we shall report briefly on a novel energy density functional (EDF) which has been derived taking into account the modification of the structure of the nucleons in the strong Lorentz scalar and vector mean-fields found in nuclei~\cite{Guichon2018,Guichon1995}. This approach has proven surprisingly successful, with results matching or exceeding in precision the results obtained with more phenomenological theoretical approaches which use typically three times the number of parameters.
\begin{table}[th!]
	\caption{Comparison of $BE$ and $R_{ch}$ residuals between the experimental values of all known even-even nuclei and those computed from 
QMC-$\pi$ III~\cite{Martinez:2020ctv}, QMC-$\pi$ II~\cite{Martinez2019}, Skyrme forces SV-min~\cite{Klup2009}, UNEDF1~\cite{Kort2012} and FRDM~\cite{Moller2016}.}
	\centering
	\begin{tabular} { c c c c c c}
	Observable		& QMC-$\pi$ III & QMC-$\pi$ II &SV-min&UNEDF1&FRDM\\\hline
        \textit{BE} (MeV)	& 1.74  &2.39 & 3.11 &2.14 &0.69\\
        \textit{R$_{ch}$} (fm$^{-3}$)&0.028 &0.026& 0.023 &0.027& not available
	\end{tabular}
\label{tab:resid}
\end{table}
\begin{figure}[tbh!]
\includegraphics[width=1.0\textwidth]{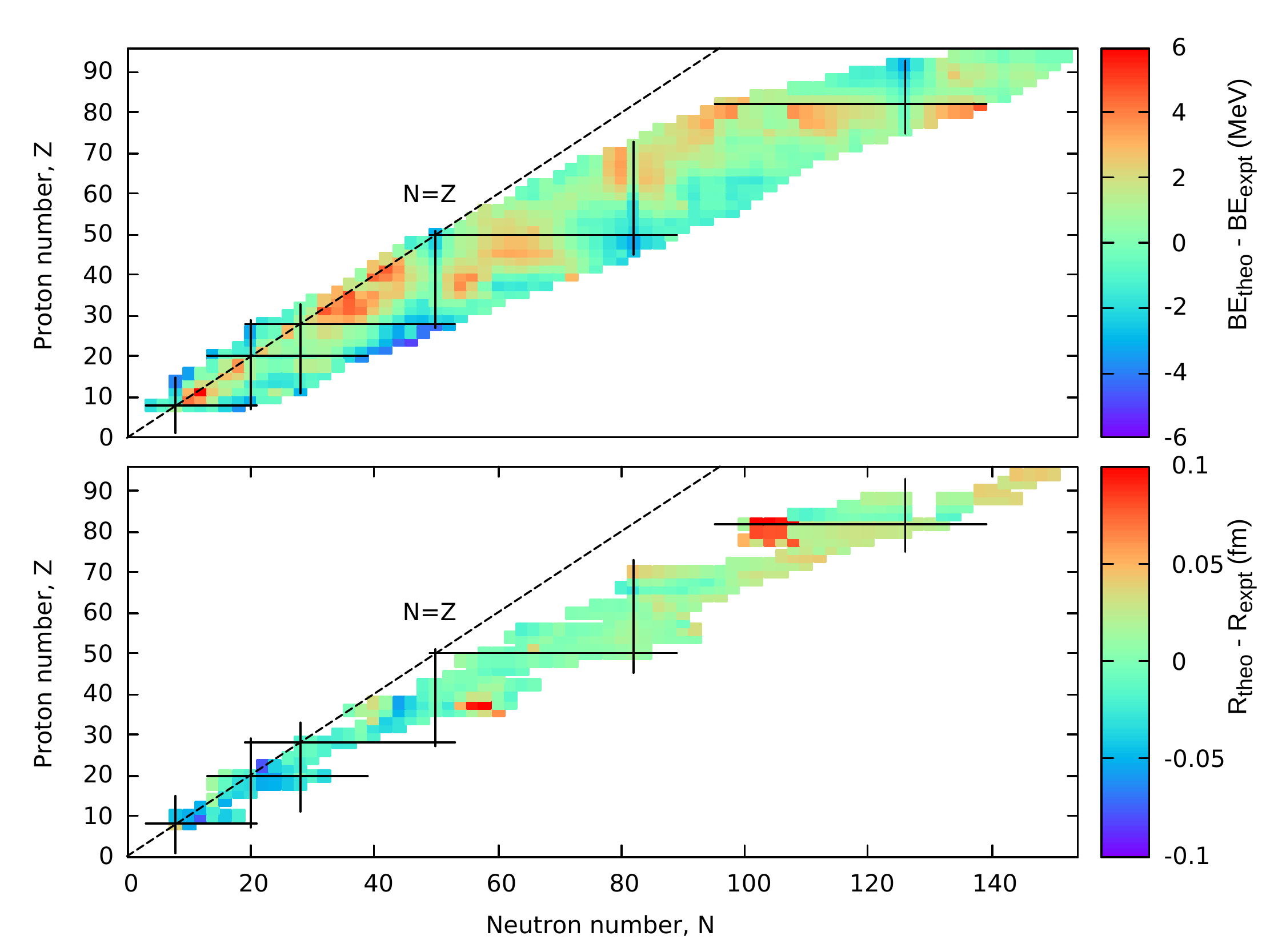}
\caption{$BE$ and $R_{ch}$ residuals for known even-even nuclei with $Z < 96$ computed from the QMC-$\pi$ III functional. Atomic mass data used to compute the binding energy residuals are taken from Ref.~\cite{Wang2017} and $rms$ charge radii data are from Ref.~\cite{Angeli2013}. Nuclei with magic numbers are indicated by solid lines and symmetric nuclei ($Z=N$) are shown in a dashed line.}
\label{fig:resid}
\end{figure}

Why does the number of parameters matter? It is well known that with five parameters Gauss suggested that he could describe an elephant. What could he have done with fifteen or more? In the present context the issue is not what one can fit but what can be predicted. The danger is that with an excess of parameters, unconstrained by physical insight, the capacity to extrapolate to new regions may not be as reliable as one would like. Fine tuning in the known region may lead to lack of control or divergence as one moves to the unknown. On the other hand, if a very small number of parameters describes the data equally well, one might suspect that it is the physics, rather than the fit routine that matters. This in turn gives one more confidence in moving to unknown regions. These considerations are especially important as we explore new regions of nuclear structure, from proton and neutron drip lines to yet to be discovered super-heavy nuclei.
\begin{figure}[tbh!]
\includegraphics[width=1.0\textwidth]{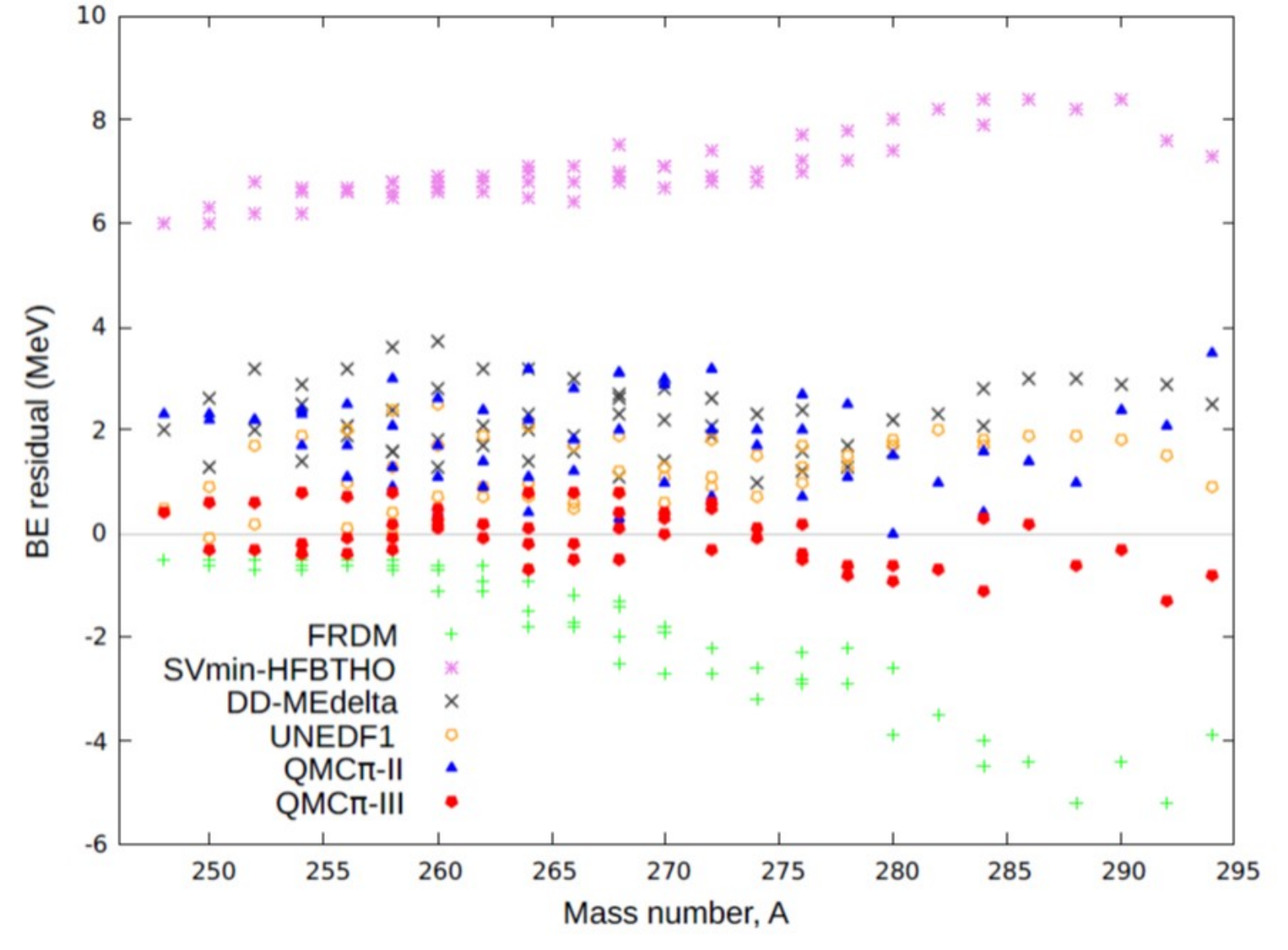}
\caption{Predictions for the binding energies of the known super-heavy nuclei from a number of models. It is remarkable that QMC-$\pi$ III, which was not fit to nuclear properties in this region, produces by far the best description of these nuclei. The rms deviation of theory from experiment is a mere 0.03\%.
}
\label{fig:SHE-Picture1}
\end{figure}

Full details of the latest EDF derived within the quark-meson coupling (QMC) framework are given in 
Ref.~\cite{Martinez:2020ctv}. This includes all of the usual components of typical Skyrme forces plus tensor and pairing interactions, all expressed in terms of just 5 QMC parameters: the sigma mass and cubic self-coupling as well as the coupling constants of the $\sigma \, , \omega$ and $\rho$ mesons to the light quarks. These parameters are chosen to reproduce nuclear matter properties at the saturation density of symmetric nuclear matter as well as the binding energies and (where known) the charge radii of 70 semi-magic nuclei with $Z \leq 94$; a total of 129 data points. The search is carried out using the derivative-free algorithm POUNDeRS~\cite{Kort2010}. The cubic $\sigma$ self-coupling is introduced to lower the nuclear incompressibility and reproduce the excitation energies of the giant monopole resonances~\cite{Martinez2019}.  
Reference~\cite{Martinez:2020ctv} also details the uncertainties in the five model  parameters, as well as the degree of correlation between them.
\begin{figure}[tbh!]
\includegraphics[width=1.0\textwidth]{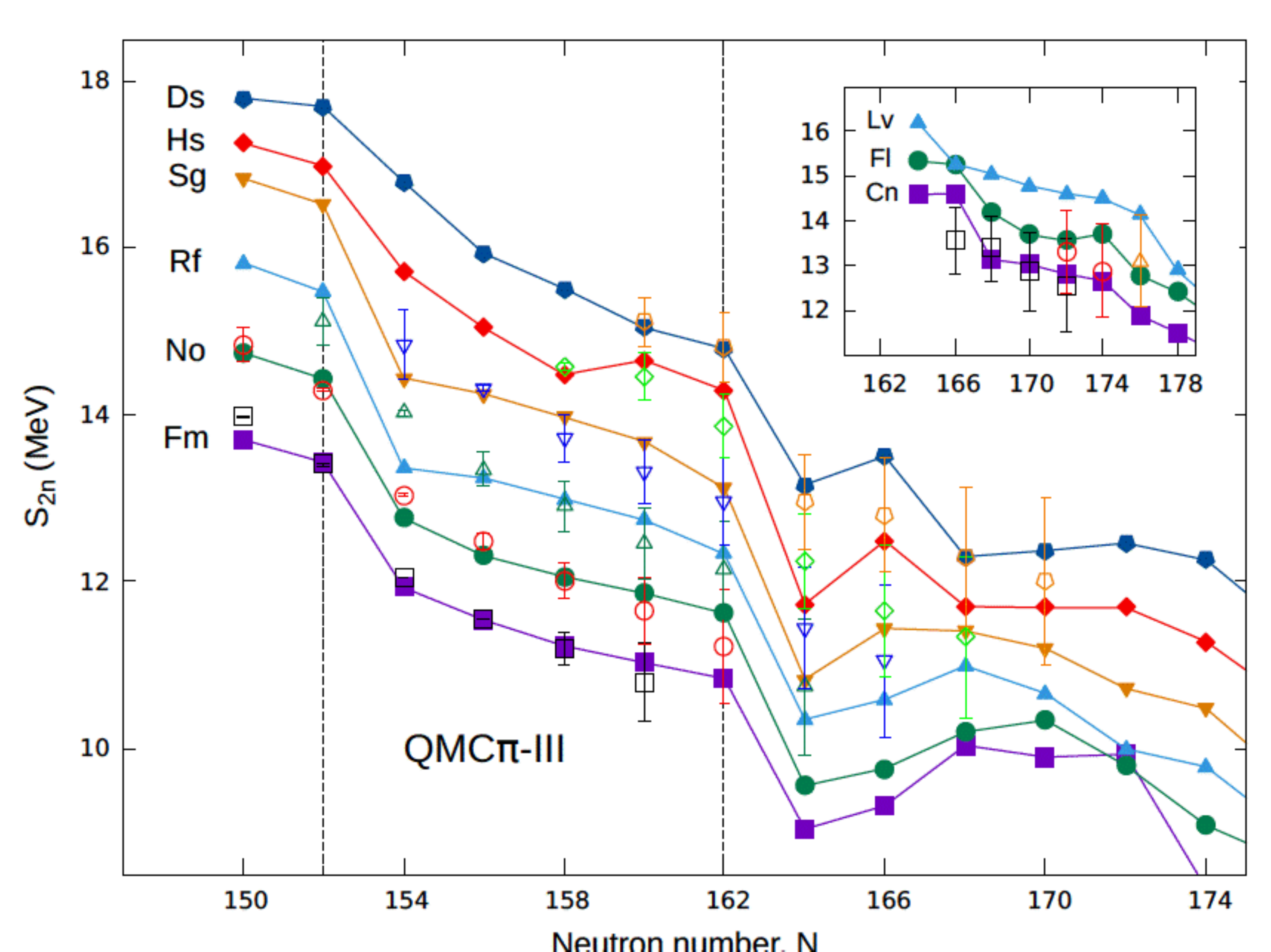}
\caption{Two-neutron separation energies plotted versus neutron
number. Results from QMC-$\pi$ III are shown as filled symbols
and connected by lines while the experimental data with errors~\cite{Wang2017} 
are shown as empty symbols.
Subshell closures at N = 152 and N = 162 are indicated with dashed
lines. Inset shows the S$_{2n}$ values plotted against $N$ for superheavy nuclei with
$Z \leq 112$.
}
\label{fig6.6}
\end{figure}
\begin{figure}[tbh!]
\includegraphics[width=1.0\textwidth]{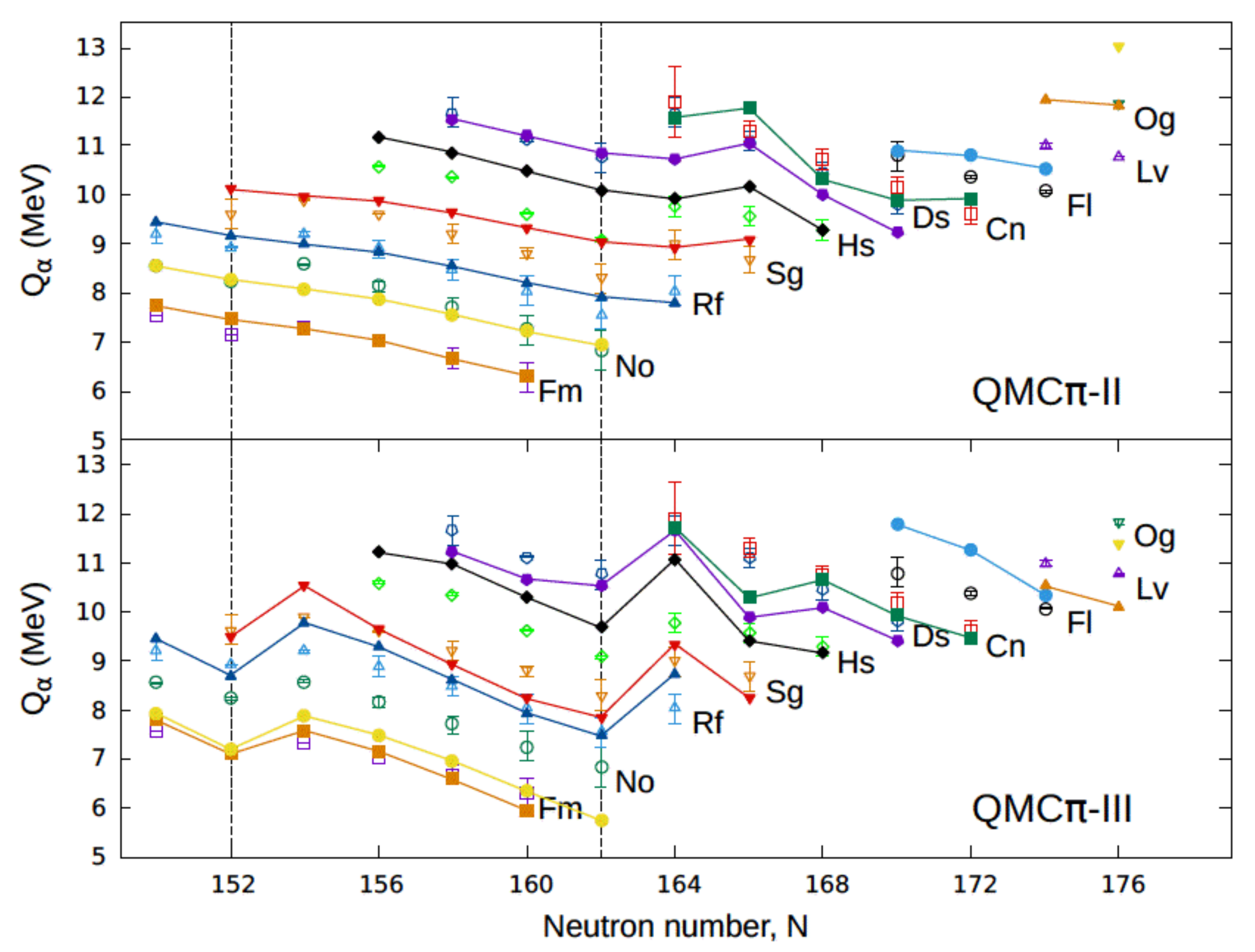}
\caption{Results for Q$_\alpha$ for nuclei with $Z \leq 100$ are shown versus neutron
number $N$. Results from QMC-$\pi$ III are shown as filled
symbols connected by lines, while the experimental data~\cite{Wang2017}
are shown as empty symbols with vertical
errorbars. Subshell closures at $N = 152$ and $N = 162$ are indicated
with dashed lines.
}
\label{fig6.9}
\end{figure}

Table~\ref{tab:resid} compares the rms deviation in the total binding energy and the rms deviation in charge radii for the two latest stages of devlopment of the QMC EDF, in comparison with other well-known models. It is clear that the QMC-$\pi$ III EDF, with just five physically motivated parameters, does as well as, or better than, the other models which all employ consioderably more parameters. This is also illustrated in Fig.~\ref{fig:resid}.
Table~\ref{tab:resid} compares the rms deviation in the total binding energy and the rms deviation in charge radiuu for the two latest stages of devlopment of the QMC EDF, in comparison with other well-known models. It is clear that the QMC-$\pi$ III EDF, with just five physically motivated parameters, does as well as, or better than, the other models which all employ considerably more parameters. This is also illustrated in Fig.~\ref{fig:resid}.

At the present time there is a great deal of interest in the existence and properties of super-heavy nuclei, that is nuclei with $Z \geq 100$. We have recently updated the first study of those 
nuclei~\cite{Martinez:2020ctv} within the QMC model. The results are shown 
in Fig.~\ref{fig:SHE-Picture1}, along with the predictions of a number of other models. With an rms deviation of just 0.03\%, the QMC-$\pi$ III EDF is far more accurate than any other. This provides a compelling case for using this model to explore the existence and properties of as yet unknown, heavier nuclei.

Another important indication of the quality of calculation of binding energies of superheavy nuclei are the two-neutron separation energies S$_{2n}$ and Q$_\alpha$ values, which are rather sensitive to subshell closures and are more important for guiding experiments than individual biding energies. Subshell closures point to a possible stability regions and the Q$_\alpha$ values are directly related to half-lives of $\alpha$ decay along decay chains, helping to identify their original parent nucleus. As an example, a change in Q$_\alpha$ by 1 MeV in a nucleus with Z = 118 would make a difference in half-life of three orders of magnitude~\cite{Stone2019}, leading to a very different decay path. It is therefore desirable to aim for predictions of Q$_\alpha$ that are as accurate as possible in order to provide a useful guide for experiment. 

 In Figs.~\ref{fig6.6} and \ref{fig6.9} we demonstrate a clear improvement using the 
QMC-$\pi$ III EDF as compared to the previous version of the model in predicting the sub-shell closure at N=152, which was predicted by the FDRM model~\cite{Moller2016} and confirmed 
experimentally~\cite{Block:2015vng}. 

Given the success of QMC-$\pi$ III across the periodic table, but especially for superheavy nuclei, the next step will be to explore the predictions of the model for as yet undiscovered nuclei beyond $z = 118$.

\newpage

\section*{Acknowledgments}
Pierre Guichon and Jirina Stone would like to acknowledge the hospitality of the CSSM at the University of Adelaide.

\section*{References}


\bibliography{Paper1} 

\end{document}